\documentclass[journal=jacsat,manuscript=article]{achemso}

\usepackage{chemformula}
\usepackage[T1]{fontenc}
\usepackage{graphicx}
\usepackage{amssymb}
\usepackage{xspace}
\usepackage{longtable}
\usepackage{soul}
\usepackage{epstopdf}
\usepackage{array}
\usepackage{tabularx}
\usepackage{physics}
\usepackage{filecontents}
\usepackage{float}
\usepackage{physics}
\usepackage[utf8]{inputenc}
\usepackage{svg}

\usepackage[colorlinks = true,linkcolor = blue,urlcolor  = blue,citecolor = blue,anchorcolor = blue]{hyperref}

\newcommand{\HoWF}{$[$Ho(W$_5$O$_{18}$)$_2]^{9-}$\xspace}
\newcommand{\HoWEn}{Na$_{9}$[Ho(W$_5$O$_{18}$)$_2$]$\cdot$35H$_2$O\xspace}

\title{Spectroscopic analysis of vibronic relaxation pathways in molecular spin qubit \HoWF: sparse spectra are key}
\author{Avery L. Blockmon}
\affiliation{These authors contributed equally to this work}
\alsoaffiliation{Department of Chemistry, University of Tennessee, Knoxville, Tennessee 37996, USA}

\author{Aman Ullah}
\affiliation{These authors contributed equally to this work}
\alsoaffiliation{Instituto de Ciencia Molecular, Universitat de Valencia, Paterna 46980, Spain}

\author{Kendall D. Hughey}
\affiliation{Department of Chemistry, University of Tennessee, Knoxville, Tennessee 37996, USA}

\author{Yan Duan}
\affiliation {Instituto de Ciencia Molecular, Universitat de Valencia, Paterna 46980, Spain}

\author{Kenneth R. O'Neal}
\affiliation{Department of Chemistry, University of Tennessee, Knoxville, Tennessee 37996, USA}

\author{Mykhaylo Ozerov}
\affiliation{National High Magnetic Field Laboratory, Tallahassee, Florida, 32310, USA}

\author{José J. Baldoví}
\affiliation {Instituto de Ciencia Molecular, Universitat de Valencia, Paterna 46980, Spain}

\author{Juan Aragó}
\affiliation {Instituto de Ciencia Molecular, Universitat de Valencia, Paterna 46980, Spain}

\author{Alejandro Gaita-Ari\~no}
\affiliation {Instituto de Ciencia Molecular, Universitat de Valencia, Paterna 46980, Spain}
\email{gaita@uv.es}

\author{Eugenio Coronado}
\affiliation {Instituto de Ciencia Molecular, Universitat de Valencia, Paterna 46980, Spain}

\author{Janice L. Musfeldt}
\affiliation{Department of Chemistry, University of Tennessee, Knoxville, Tennessee 37996, USA}
\alsoaffiliation{Department of Physics, University of Tennessee, Knoxville, Tennessee 37996, USA}
\email{musfeldt@utk.edu}

\date{\today}

\begin{document}

\begin{abstract}

Molecular vibrations play a key role in magnetic relaxation processes of molecular spin qubits as they couple to spin states, leading to the loss of quantum information. Direct experimental determination of vibronic coupling is crucial to understand and control the spin dynamics of these nano-objects, which represent the limit of miniaturization for quantum devices. Herein, we measure the vibrational properties of the molecular spin qubit \HoWEn
by means of magneto-infrared spectroscopy. Our results allow us to unravel the vibrational decoherence pathways in combination with $ab$ $initio$ calculations including vibronic coupling.
We observe field-induced spectral changes near 63 and 370 cm$^{-1}$ that are modeled in terms of $f$-manifold crystal field excitations activated by odd-symmetry vibrations. The overall extent of vibronic coupling in this system is limited by a transparency window in the phonon density of states that acts to keep the intramolecular vibrations and $M_J$ levels apart. These findings advance the understanding of vibronic coupling in molecular magnets, place significant constraints on the pattern of crystal field levels in these systems,
and provide a strategy for designing molecular spin qubits with improved coherence lifetimes.

\vspace{0.15in}


\end{abstract}

\maketitle

%%%%%%%%%%%%%%%%%%%%%%%%%%%%%%%%%%%%
\section*{Introduction}
%%%%%%%%%%%%%%%%%%%%%%%%%%%%%%%%%%%%

Quantum two-level systems based on spin states known as “spin qubits” are promising building blocks for the development of quantum technologies. In contrast with classical bits that are defined by two states “0” or “1”, qubits exploit quantum behavior by allowing quantum superposition between the basis states. Among different physical platforms, chemistry contributes to this effort %fundamentally
via the study of spin states in magnetic molecules, both in the form of molecular spin qubits and molecular nanomagnets. In this context, mononuclear lanthanide complexes provide ideal %laboratories
settings that represent the limit of miniaturization for quantum and classical magnetic memories, respectively.\cite{coronado2020molecular, gaita2019molecular, atzori2019second}
%I'd rather start the introduction with this approach.
%Single molecule magnets incorporating heavy centers are of great fundamental and practical importance.\cite{Rinehart2011, Ungur2016,Harriman2017,Long2018} They differ from 3$d$ transition metal-containing counterparts,\cite{Hoeke2014, Nandy2020} in terms of diffused vs. localized orbitals and the nature of chemical bonding, the relative importance of electron correlation vs. spin-orbit coupling, and the energy of the crystal field levels - all of which impact properties and coupling.\cite{Pedersen2010, Wang2011,ONeal2019}
%%orbital localization/diffusivity and  chemical bonding, spin-orbit coupling, the blocking temperature, the crystal field levels, as well as ultra-low frequency vibrational modes. These factors in turn impact the properties.
%%The addition of heavy centers is also opportunity to increase  magnetoelectric, vibronic, and spin-phonon coupling.
Hundreds of such systems have been characterized so far in an attempt to understand and control their physical properties.\cite{Woodruff2013} A judicious chemical design exploiting magnetic anisotropy and optimizing the molecular structure versus spin-vibrational coupling led to the observation of magnetic bistability up to 80 K and an effective energy barrier of 2217 K in the molecular nanomagnet [(Cp$^{iPr5}$)Dy(Cp$^{*}$)]$^+$\cite{Guo2018}. Regarding molecular spin qubits, prominent examples include: two vanadium(IV) complexes, namely (i) [V(C$_8$S$_8$)$_3$]$^{2-}$, which shows a record phase memory time $T_{\rm M}$ of 0.7 ms\cite{zadrozny2015millisecond} and (ii) VOPc, with coherence times up to 1 $\mu$s at room temperature\cite{atzori2016room}; (iii) a Cu$^{2+}$ complex, [P(Ph$_4$)]$_2$[Cu(mnt$_2$)] (mnt$^{2-}$=maleonitriledithiolate), that preserves coherence up to room temperature\cite{bader2014room} when diluted in Ni$^{2+}$; and (iv) a Ln$^{3+}$-based molecular nanomagnet, [Ho(W$_5$O$_{18}$)$_2$]$^{9-}$, whose spin qubit dynamics are protected against magnetic noise at optimal operating points known as atomic clock transitions.\cite{Shiddiq2016}.
%include: (i) [(Cp$^{iPr5}$)Dy(Cp$^{*}$)]$^+$
%(Cp$^{iPr5}$ = penta-iso-propylcyclopentadienyl; Cp$^{*}$ = pentamethylcyclopentadienyl)
%with a record-high 80 K blocking temperature,  effective energy barrier,  slow magnetic relaxation, and spin-phonon coupling, \cite{Guo2018} (ii) $R,R-$[Zn(OAc)(L)Yb(NO$_3$)$_2$]
%$R,R$-H$_2$L = [$6,6'$-((1E,1'E)-(((1R,2R)-1,2-diphenylethane-1,2-diyl) bis (azaneylylidene)) bis (methaneylylidene)) bis(2-methoxyphenol)]
%with room temperature magnetoelastic and magnetoelectric coupling as well %as ferroelectricity,\cite{Long2020} and (iii) Na[Dy(DOTA)(H$_2$O)]·4H$_2$O that displays strong magnetic anisotropy, spin-orbit coupling, and photoluminescence.\cite{Car2011, Cucinotta2012}
%H$_4$DOTA = 1,4,7,10-tetraazacyclododecane-N,N',N'',N'''-tetraacetic acid.
%Applications in high-density memory storage, as contrast agents, and in molecular spintronics and quantum computing.\cite{Mills2011,Woodruff2013,Coronado2019,Gonzalez2019}
%
In this work, we will focus on the latter, investigating the vibronic relaxation pathways\cite{Hackermuller2004,Chirolli2008, Graham2014,Goodwin2017} that govern magnetic relaxation. Thus far, these pathways have been largely under-explored. In particular, while transitions between electron-nuclear spin states are protected from quantum decoherence at the atomic clock transitions, this protection of the quantum information is only from magnetic noise. Instead, the system suffers from a reduction of relaxation time $T_2$ by a factor of $1/2$ between 5 and 7 K due to thermal noise, which limits coherence.\cite{Shiddiq2016}

\begin{figure}[tbh!]
\centering
\includegraphics[width=2.0in]{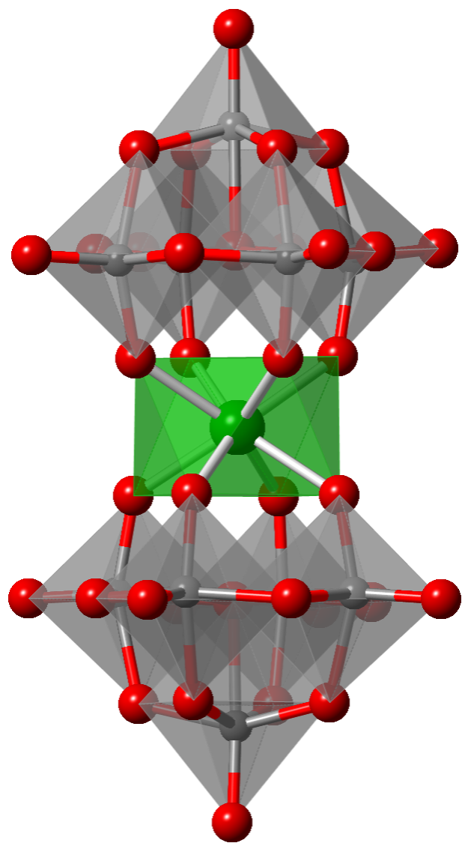}\\
\caption{Structure of the molecular anion [Ho(W$_5$O$_{18}$)$_2$]$^{9-}$. Color code: Ho (green), W (gray) and O (red).}
\label{HoW10molecule}
\end{figure}

Single molecule magnet behavior in Na$_{9}$[Ho(W$_5$O$_{18}$)$_2$]$\cdot$35H$_2$O arises from the magnetic anisotropy of a Ho$^{3+}$ ion ($J$ = 8) encapsulated by two polyoxometalate  moieties in an eight-fold oxygen coordination environment with local slightly distorted $D_{4d}$ symmetry (Fig. \ref{HoW10molecule}). \cite{Shiozaki1996,AlDamen2009,Gu2018} This assignment stabilizes a $M_J$ = $\pm$4 ground state. In the crystal, the  [Ho(W$_5$O$_{18}$)$_2$]$^{9-}$ anions are held together by interactions with Na$^{+}$ ions and free water molecules of crystallization with a space group $P{\bar 1}$.\cite{AlDamen2008} The magnetic properties of Na$_{9}$[Ho(W$_5$O$_{18}$)$_2$]$\cdot$35H$_2$O have attracted significant attention. %been extensively investigated.
The out-of-phase $ac$ magnetic susceptibility displays a maximum near 5 K, \cite{AlDamen2009}
and magnetization shows a slow relaxation \cite{AlDamen2009}, one of the key signatures of single molecule magnets.\cite{Li2006}
Electron paramagnetic resonance sports an eight line spectrum due to the ground $M_J$ level hyperfine coupling to the $I = 7/2$ nuclear spin along with a tunneling gap of approximately 9 GHz (0.3 cm$^{-1}$).\cite{Ghosh2012}
Two-pulse electron spin echo measurements at its optimal working points reveal spin coherence times up to $T_2$=8.4 $\mu$s at 5 K when diluted to a 1\% concentration in the isostructural diamagnetic matrix of its Y$^{+3}$-based analog, Na$_{9}$[Ho$_{0.01}$Y$_{0.99}$(W$_5$O$_{18}$)$_2$]$\cdot$35H$_2$O. The high dilution limit for $T_2$ in other molecular spin qubits is usually achieved at concentrations around 0.01\%.\cite{Shiddiq2016} Recent results operating at these special points reveal the possibility of quantum magnetoelectric coupling, where voltage control of the crystal field levels can be used to manipulate the spin qubit states.
%controlling its spin states via electrical control of its crystal field.
\cite{liu2020quantum}

Despite the importance of the crystal field levels in this system, spectroscopic information is limited to the aforementioned 0.3 cm$^{-1}$ tunneling splitting in the ground doublet determined by electron paramagnetic resonance  as well as the low energy $M_J$ levels at 41 and 49 cm$^{-1}$ measured by inelastic neutron scattering by Vonci, {\it et al.}\cite{Vonci2017}
A number of authors have attempted to locate the position and determine the order of the rest of the crystal field energy levels in
Na$_{9}$[Ho(W$_5$O$_{18}$)$_2$]$\cdot$35H$_2$O.\cite{Baldovi2014, Vonci2017} While it is very much accepted that the seventeen different $M_J$ levels reside between 0.3 and 400 cm$^{-1}$,\cite{Baldovi2014, Vonci2017} the exact pattern and order has been elusive due to the anisotropy of the system along with coupling to the vibrational levels.\cite{Vonci2017}
Moreover, Ho$^{3+}$ is a non-Kramers ion, so while at first glance, it appears that many of the crystal field levels are doubly-degenerate, there is a very small
%(0.085 cm$^{-1}$)
% giving this number seems misleading, since it distracts from the 0.3cm-1 tunneling splitting in the ground doublet, which is more important
energy scale that splits the $M_J$  levels and relieves  their apparent  degeneracies.\cite{Vonci2017}
Although there is some prior infrared spectroscopy,\cite{AlDamen2009} the work focused only on the middle infrared and, as a result, did not explore excitations below 400 cm$^{-1}$ which play a key role as they may interact with spin excitations, changes in spin state, and crystal field energy levels.\cite{Brinzari2013, Brinzari2013b}
At the same time, the vibrational features in Na$_{9}$[Ho(W$_5$O$_{18}$)$_2$]$\cdot$35H$_2$O are unassigned,\cite{AlDamen2009} making it difficult to % use the results to
test the order of the Ho$^{3+}$ crystal field excitations or model coupling processes.
This is important because, as a candidate qubit material, it is crucial to unravel all decoherence mechanisms that emanate from lossy responses in this frequency range.\cite{Hartle2011,Ballmann2012}
%The Ballmann and Hartle references deal with vibronic coupling being a decoherence mechanism

In order to explore the dynamics of a model molecular spin qubit system, we combined far infrared spectroscopy and magneto-infrared techniques with complementary lattice dynamics and vibronic coupling models to reveal the low energy excitations in Na$_{9}$[Ho(W$_5$O$_{18}$)$_2$]$\cdot$35H$_2$O.
Strong magneto-infrared effects are observed near 370 and 63 cm$^{-1}$, which are due to activation of crystal field excitations by molecular vibrations of appropriate symmetry.
%
%Specifically, the $M_J$ = $\pm$7 crystal field levels centered at 370 cm$^{-1}$ couple to the Ho-O$_8$ rocking and stretching modes.
%The $M_J$ = $\pm$5 levels near 63 cm$^{-1}$ are activated by nearby phonons as well.
These structures place significant constraints on the position of several $M_J$ levels - including the highest energy set - establishing stringent bounds on the entire series of excitations.
Our analysis suggests that coherence  in Na$_{9}$[Ho(W$_5$O$_{18}$)$_2$]$\cdot$35H$_2$O benefits strongly from the limited frequency overlap between the crystal field levels and the phonon manifold. We demonstrate that the limited frequency overlap is due to a large transparency window or “hole” in the phonon density of states.
%which renders many of the vibronic pathways in this system ineffective in terms of generating decoherence. %This is because fewer level crossings reduce loss.
%relatively large separation between phonons and the crystal field levels.
This provides a strategy for the chemical design of molecular nanomagnets and spin qubits in which a full separation of the electronic and vibrational excitations will be likely to entirely eliminate vibronic coupling as a decoherence mechanism.\cite{escalera2018spin,ullah2019silico, garlatti2020unveiling, yu2020spin}

%%%%%%%%%%%%%%%%%%%%%%%%%%%%%%%
\section*{Results and discussion}
%%%%%%%%%%%%%%%%%%%%%%%%%%%%%%%

%%%%%%%%%%%%%%%%%%%%%%%%%%%%%%%%%%
\subsection*{Assigning the low energy excitations in Na$_{9}$[Ho(W$_5$O$_{18}$)$_2$]$\cdot$35H$_2$O}
%\subsection*{Far-IR magnetospectroscopy studies}
%%%%%%%%%%%%%%%%%%%%%%%%%%%%%%%%%%

\begin{figure}[b!]
\centering
\includegraphics[width=4.2in]{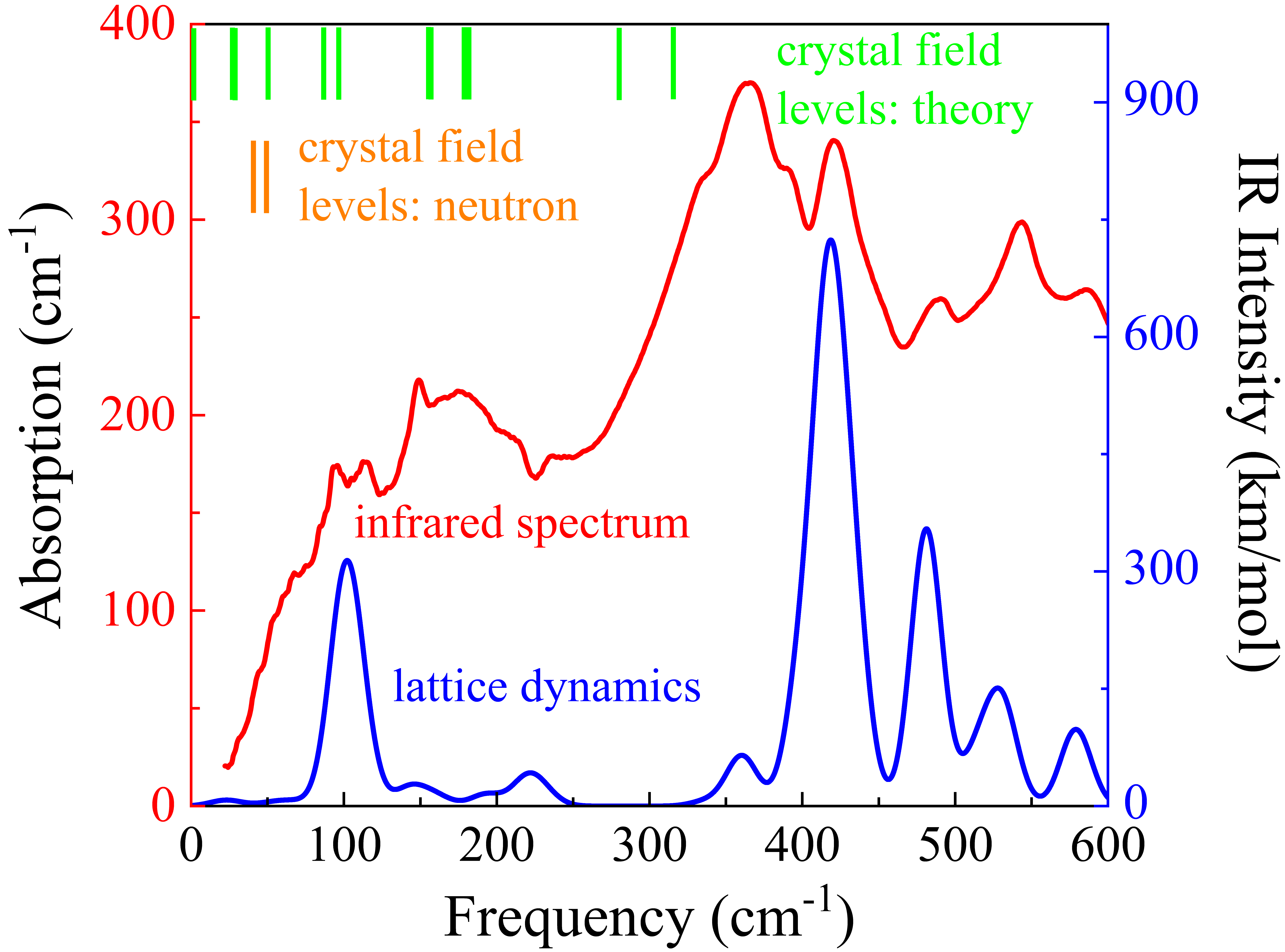}\\
\caption{Far infrared spectrum of Na$_{9}$[Ho(W$_5$O$_{18}$)$_2$]$\cdot$35H$_2$O at room temperature compared with calculated lattice dynamics and crystal field levels as well as neutron scattering data from Ref \citenum{Vonci2017}.}
\label{IRspec}
\end{figure}

Figure \ref{IRspec} summarizes the far infrared response of Na$_{9}$[Ho(W$_5$O$_{18}$)$_2$]$\cdot$35H$_2$O.
As a reminder, we focused our efforts on the low frequency regime  from 25 - 600 cm$^{-1}$  because (i) the far infrared contains key vibrations involving the Ho$^{3+}$  center as well as crystal field levels, and (ii) only the middle infrared was studied in prior work.\cite{AlDamen2009}
Figure \ref{IRspec} includes a number of important energy levels, predictions, and simulations  for comparison. Many of these excitations are vibrational in nature, and with the aid of our lattice dynamics calculations, we can assign the peaks and model the overall spectrum.
For instance, we assign the dominant peak at 365 cm$^{-1}$ as a rocking motion of HoO$_4$. This rocking motion is accompanied by minor distortions of the (W$_{5}$O$_{18}$)$^{6-}$ cages. The theoretically predicted vibrational response (appropriately broadened) is shown in blue for comparison. The overall agreement with the measured spectrum is satisfactory. Interestingly, our lattice dynamics calculations predict a broad range - from approximately 230 to 330 cm$^{-1}$ - where the fundamental excitations of the lattice are absent. While less apparent than in the theoretical spectrum, this characteristic of mode clusters below 230 cm$^{-1}$ and above 330 cm$^{-1}$ is evident in the experimental spectrum as well. The absence of spectral features in this region is a consequence of a “hole” in the phonon density of states. The implications of this structure will be discussed below.  Complete vibrational mode assignments are available in Table S1 of the Supplemental Information.

$f$-block elements like Ho$^{3+}$ also display crystal field levels in the far infrared.\cite{Marx2014, Kumar2016, Ungur2017}
The exact position of the $f$-manifold levels varies significantly depending upon the details of the crystal field environment.\cite{Newman1971, Goodwin2017, Hallmen2018}
Previous authors located the $M_J$ = $\pm$ 3 and $\pm$ 5 levels in Na$_{9}$[Ho(W$_5$O$_{18}$)$_2$]$\cdot$35H$_2$O via inelastic neutron scattering.\cite{Vonci2017}
These energies are shown in Fig. \ref{IRspec} in orange
and help to define a lower limit to the crystal field manifold.
%
%In this work, we computed the full set of Ho$^{3+}$ crystal field levels in this system using the multireference Complete Active Self-Consistent Field Spin-Orbit (CASSCF-SO) method (implemented by OpenMOLCAS \cite{fdez2019openmolcas}) with the combined effect of the crystal field and the spin-orbit coupling calculated using SINGLE-ANISO module\cite{ungur2017ab}.
%
As part of this work, we computed the full set of Ho$^{3+}$ crystal field levels for this system. They are included in Fig. \ref{IRspec} in green for comparison. Our predictions for the position of the $M_J$ = $\pm$3 and $\pm$5 levels are consistent with the neutron scattering results.\cite{Vonci2017}  The upper  $M_J$ levels obtained from our model are also in close proximity %coincident
to various molecular vibrations involving the central holmium atom and coordinated oxygen atoms. %the HoO$_8$ rocking and stretching modes in Na$_{9}$[Ho(W$_5$O$_{18}$)$_2$]$\cdot$35H$_2$O.
These features are on the leading edge of the upper cluster of modes above the 230 - 330 cm$^{-1}$ transparency window or “gap” in the infrared spectrum.

\begin{figure}[tbh]
\centering
\includegraphics[width=6.5in]{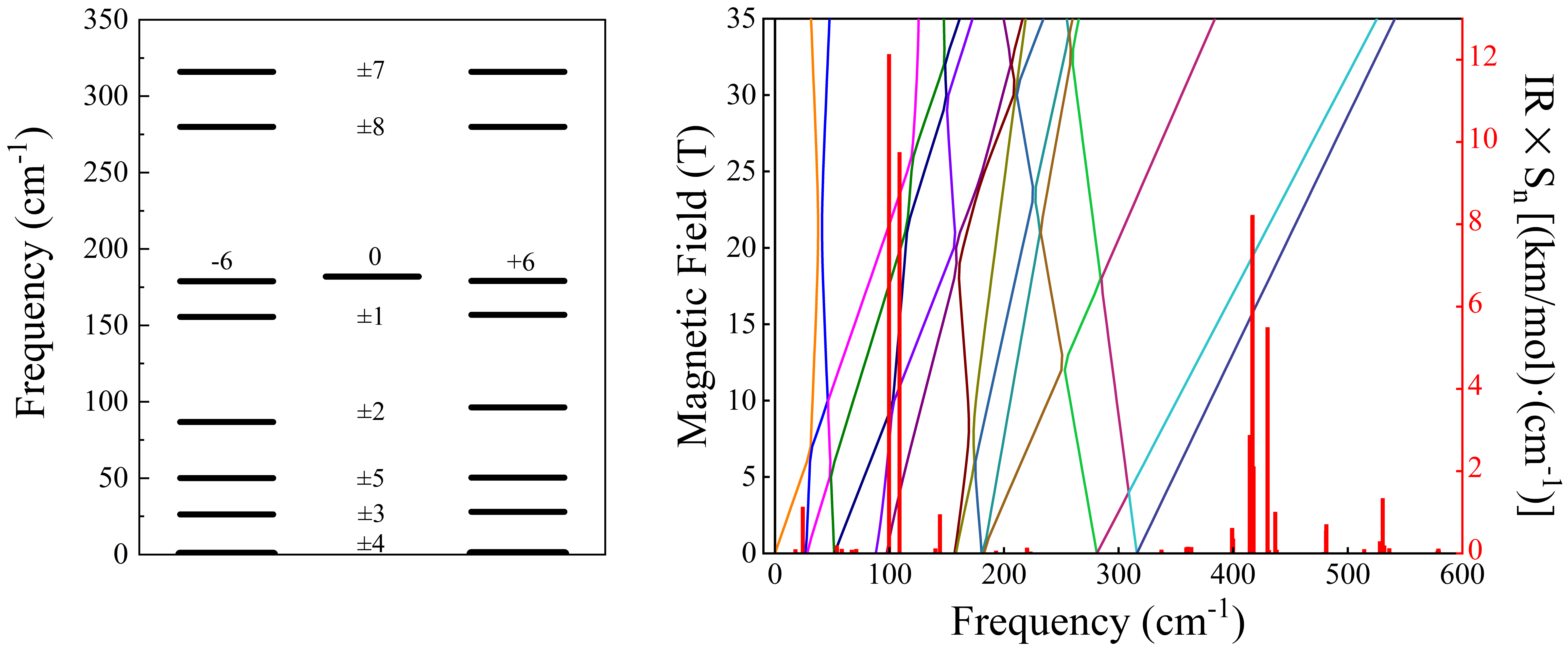}\\
\caption{Left: Calculated crystal field levels of [Ho(W$_5$O$_{18}$)$_2$]$^{9-}$. Note that the energies are normalized to the ground $M_J$ state. Right: Double y-axis plot of calculated crystal field levels with magnetic field and vibrational mode intensity times vibronic coupling (S$_n$).}
\label{Fig:CFvsB}
\end{figure}

Figure \ref{Fig:CFvsB} shows the calculated crystal field levels in greater detail along with their evolution under applied magnetic field along the long molecular axis (see the evolution for other directions of the magnetic field in the Supplementary Information). As expected, the Zeeman effect between 0 and 35 T widens the total spread of the crystal field levels, with each level acquiring a slope proportional to its expectation value $<M_J>$. The slopes becoming non-linear as a result of either avoided crossings among crystal field levels or transverse magnetic fields.

 Then we measured the far infrared response of Na$_{9}$[Ho(W$_5$O$_{18}$)$_2$]$\cdot$35H$_2$O as a function of temperature. As shown in Fig. S2 of the Supplemental Information, the lattice is nearly rigid.
 These findings were verified by complementary measurements of the non-magnetic analog Na$_{9}$[Y(W$_5$O$_{18}$)$_2$]$\cdot$35H$_2$O
which reveals a similar lack of temperature effects. Further, the Ho and Y analogs have identical transparency windows in the infrared response (Fig. S1 of the Supplemental Information). The phonon density of states is therefore not radically different than in the diluted case represented by Na$_{9}$[Ho$_{0.01}$Y$_{0.99}$(W$_5$O$_{18}$)$_2$]$\cdot$35H$_2$O.\cite{Shiddiq2016}
 %When tracking peak position vs. temperature, e
 % where every feature blue shifts with decreasing temperature.
 Although there are uncoordinated waters in the crystal structure, the slight blue shifting of peaks demonstrates that hydrogen bonding does not strengthen significantly at low temperature. \cite{Jones2001}

%%%%%%%%%%%%%%%%%%%%%%%%%%%%%%%%%%%%
\subsection*{Magneto-infrared response of Na$_{9}$[Ho(W$_5$O$_{18}$)$_2$]$\cdot$35H$_2$O}
%%%%%%%%%%%%%%%%%%%%%%%%%%%%%%%%%%%%

\begin{figure}[tbh!]
\flushright
\includegraphics[width=6in]{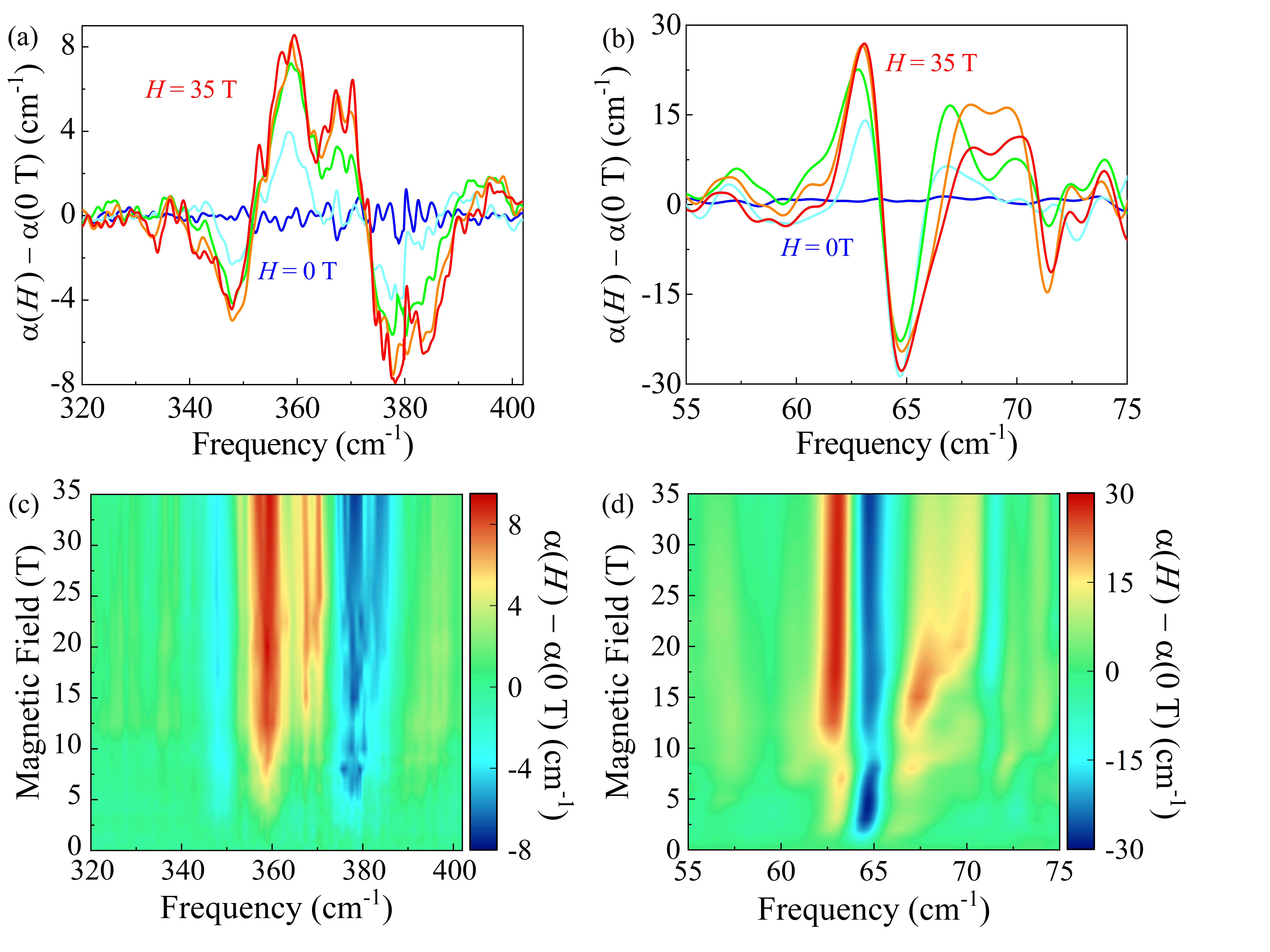}
\caption{(a,b) Magneto-infrared response of the features centered at 370 and 63 cm$^{-1}$, respectively, at 0, 5, 12.5, 20 and 35 T. The data are shown as absorption differences: $\Delta\alpha({\omega},H) = \alpha({\omega},H) - \alpha({\omega},H$ = 0~T).
(c,d) Contour plots of the absorption difference spectra from panels (a,b) over the full range of magnetic fields. Scale bars are included.}\label{Multipanel}
\end{figure}

Figure \ref{Multipanel} (a,b) displays the magneto-infrared response of Na$_{9}$[Ho(W$_5$O$_{18}$)$_2$]$\cdot$35H$_2$O at 4.2 K.
The spectra are in the form of absorption differences,
 $\Delta\alpha(H)  = \alpha(H) - \alpha$($H$ = 0~T), which serves to highlight field-induced changes. The latter are not always apparent in the absolute absorption.
There are two primary structures that develop systematically under applied  field: (i)  a set of sharp features centered near 63 cm$^{-1}$,  and (ii) several broader, more complex
structures between 335 and 400 cm$^{-1}$. Contour plots also show the development of the absorption differences with increasing magnetic field (Fig. \ref{Multipanel} c,d).
%In other words, contrast as represented by the absorption difference increases with  field.
For comparison purposes, we performed similar measurements on the non-magnetic analog Na$_{9}$[Y(W$_5$O$_{18}$)$_2$]$\cdot$35H$_2$O. No field-induced spectral changes were observed - as anticipated.

We assign contrast in the absorption difference spectra of Na$_{9}$[Ho(W$_5$O$_{18}$)$_2$]$\cdot$35H$_2$O based upon the position of the intramolecular vibrations and $f$-manifold crystal field levels  discussed in  Fig. \ref{IRspec}.
The magneto-infrared  features centered near 360  cm$^{-1}$ are in close proximity to several different vibrational modes including  HoO$_{8}$ rocking and stretching modes - all of which take place in the center of the cage.
Therefore, at first glance, it may seem that the magneto-infrared response in this frequency region may be due to spin-phonon coupling. However, the lack of magnetic order (or change in magnetic order) in the system argues against this interpretation.\cite{Casto2015}
Other candidate excitations include crystal field levels. The highest frequency set, $M_J$ = $\pm$7 are calculated by ourselves and others\cite{Vonci2017}  to be in this vicinity.
This suggests that vibronic coupling - in which the crystal field levels move with applied magnetic field and are amplified by nearby vibrational modes - may be responsible for the magneto-infrared contrast.\cite{Brinzari2013b, Sethi2019, Tang2020} We test both models in this work and establish the vibronic coupling model below. Thus, the assignment of the $M_J$ = $\pm$7 levels interacting with nearby phonons serves as an upper bound for the entire $f$-manifold of Ho$^{3+}$ excitations.

A similar mechanism can explain the contrast  centered near 63 cm$^{-1}$ (Fig. \ref{Multipanel}b). % although the details are less obvious.
While our dynamics calculations predict significant intensity for localized vibrational modes starting at 100 cm$^{-1}$ (Fig. \ref{IRspec}),  Na$_{9}$[Ho(W$_5$O$_{18}$)$_2$]$\cdot$35H$_2$O certainly has librational modes of uncoordinated H$_2$O molecules\cite{Tayal1980, Dominguez2007} as well as lattice modes involving asymmetric stretching of the HoO$_8$ along with cage tilting.  While the details depend upon the precise environment of the uncoordinated water as well as the long-range crystallographic orientation of the polyoxometalates, these vibrations can act as symmetry-breakers for the large number of $f$-manifold excitations in the vicinity. In other words, the contrast centered near 63 cm$^{-1}$ in the magneto-infrared spectra of  Na$_{9}$[Ho(W$_5$O$_{18}$)$_2$]$\cdot$35H$_2$O is due to vibrational modes activating excitations involving nearby crystal field levels -  specifically the $M_J$ = $\pm$5 levels. It is possible that the $\pm$2 levels contribute as well. This assignment is based upon our detailed calculations of the  crystal field levels (Fig. \ref{Fig:CFvsB}).
In any case, these crystal field levels also shift with applied magnetic magnetic field and can vibronically couple with appropriate nearby vibrations.

To elucidate the origin of the magneto-infrared response, we simulated the absorption spectra at different fields. In absence of vibronic coupling, the selection rules would mean a trivial simulation that fails to reproduce the experimental behavior. Indeed this would result in the sum of a field-independent purely vibrational spectrum as shown in Fig. \ref{IRspec} and a collection of strongly field-dependent but strongly forbidden transitions among crystal field levels. Instead, we employed a more realistic model to estimate the transition dipole matrix between all possible vibrational and crystal field levels that crucially considers vibronic eigenfunctions, allowing for mixing between the two kinds of excitations (see details in Supplementary Information).

\begin{figure*}[tbh!]
\centering
\includegraphics[width=6.5in]{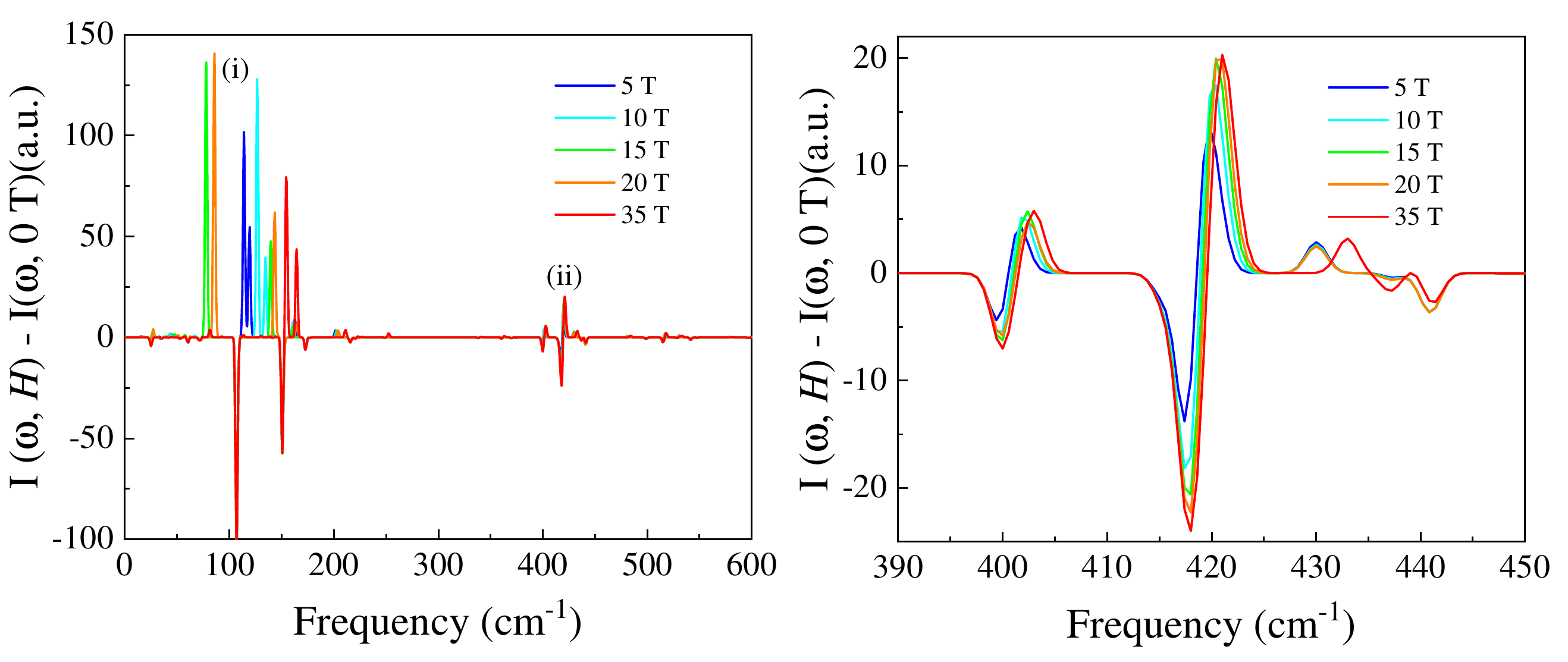}
\caption {Left: Simulated magneto-infrared spectra of [Ho(W$_5$O$_{18}$)$_2$]$^{9-}$ between 5 and 35 T. Two main regions of interest are denoted. Right: Close-up view of the magneto-infrared simulated response between 390 and 450 cm$^{-1}$.}
\label{IRAb}
\end{figure*}

We simulate the magneto-infrared response of Na$_{9}$[Ho(W$_5$O$_{18}$)$_2$]$\cdot$35H$_2$O employing this approach (Fig. \ref{IRAb}). We examined the absorption differences in spectra up to 35 T and find consistent results, with two regions of notable effects at relatively low and high frequencies. They show a wide gap in the spectrum between 200 and 390 cm$^{-1}$ devoid of any significant anti-crossings. Let us discuss first the region above that gap (emphasized in Fig. \ref{IRAb} b), which corresponds to the better determined experimental features. In this highest frequency region, the evolution of the spectra with increasing magnetic field is very simple and qualitatively matches the experimental results. In particular, the calculations reproduce the gradual saturation of the effect of the magnetic field in the absorption intensity, where the linear increase up to 10-15 T is lost and only a minor change is observed between 20 T and 35T (compare with Fig.~S4). Overall the agreement is very satisfactory, save for a rigid 40 cm$^{-1}$ blue shift from our experimental features. In the case of the low energy region, where the theoretically estimated effects are also similarly shifted to slightly higher energies, both in the experiment and in the calculations, the behavior is more complex. The region between 80 and 200 cm$^{-1}$ where the calculations predict Zeeman-like displacements of absorptions occur in the experiments by a complex, possibly noisy, pattern. This is, in fact, what we observe (Fig. S3 of the Supplemental Information).

\subsection{Developing molecular design criteria}

The main goal for the rational design of longer-lived molecular spins is decoupling spin energy levels from vibrational excitations.
Our experimental and theoretical work confirms
%Herein we present experimental and theoretical\cite{Ullah2019} confirmation
that the key towards this goal is two-fold.
%which where pointed out on purely theoretical level.\cite{Ullah2019}
First, one needs a sparse low-energy crystal field spectrum, resulting from a very strong crystal field as in dysprosocenium\cite{Guo2018}, from a bunching of crystal field levels at higher energies as in Tb bis-phthalocyaninato sandwiches\cite{ishikawa2003lanthanide} or from incorporating U as a magnetic center.\cite{escalera2020design} This can be combined with the optimization via chemical design of different coordination environments that will displace the $M_J$ levels to other regions of the spectra. Second, one needs an effectively sparse low frequency vibrational spectrum, resulting from a molecular environment for the spin that is either very simple, like in small molecules, with fewer degrees of freedom,\cite{zadrozny2015millisecond, bader2014room} or very rigid, as in fully aromatic ligands,\cite{atzori2016room} where vibrations are pushed to higher frequency or engage in exceptionally limited vibronic coupling.
The point is that by placing these features off-resonance, thermal routes to decoherence may be weakened and potentially curtailed. %reduced.

In terms of decoherence pathways in the system Na$_{9}$[Ho$_x$Y$_{(1-x)}$(W$_5$O$_{18}$)$_2$]$\cdot$35H$_2$O, our analysis allows the qualitative rationalization of the strong thermal dependence of the coherence time $T_2$ in the so-called atomic clock transitions. $T_2$ was experimentally determined to be $T_1$-limited up to concentrations of $x$ = 0.01 and down to temperatures of 5 K.\cite{Shiddiq2016} In the present work we demonstrate that there is a relatively dense population of energy levels, both crystal field and vibrational, in the region below 100 cm$^{-1}$, and these present significant vibronic coupling, as evidenced both experimentally and theoretically. This means that even a small population of these levels at relatively low temperatures is expected to have a significant detrimental effect on the coherence. In contrast, there is a limited frequency overlap between the crystal field levels and the phonon manifold in the window between 200-330 cm$^{-1}$. Should it be possible to alter the coordination to place this large ``hole'' in the phonon density of states at lower energies, it would render many of the vibronic pathways in this system ineffective in terms of generating decoherence. The transparency window in the phonon density of states renders some pathways ineffective even now.
A full separation of the electronic and vibrational excitations is likely to entirely eliminate vibronic coupling as a decoherence mechanism.

%%%%%%%%%%%%%%%%%%%%%%%%%%%%%%%%
\section*{Concluding remarks}
%%%%%%%%%%%%%%%%%%%%%%%%%%%%%%%%

We combined far infrared spectroscopy and magneto-infrared techniques with complementary lattice dynamics and vibronic coupling models to unveil the low energy excitations in Na$_{9}$[Ho(W$_5$O$_{18}$)$_2$]$\cdot$35H$_2$O and gain insight into the spin dynamics of a model qubit system.
We found strong magneto-infrared effects near 370 and 63 cm$^{-1}$ due to the activation of crystal field excitations via molecular vibrations of appropriate symmetry.
Specifically, the $M_J$ = $\pm$7 crystal field levels couple to the various HoO$_8$ rocking and stretching modes.
On the other hand, the $M_J$ = $\pm$5 levels near 63 cm$^{-1}$ (and very likely the $M_J$ = $\pm$2 levels) are activated by nearby phonons (such as H$_2$O librations and asymmetric HoO$_8$ stretching with cage tilting).
%These magneto-infrared structures place significant constraints on the position of several $M_J$ levels –including the highest energy set– and establish stringent bounds on the entire series of excitations.
%
At the same time, we report the first direct evidence for a transparency window  in the phonon density of states in a robust molecular spin qubit. Besides exploring the consequences of this ``hole'' on vibronic coupling, we use our findings to propose new design rules aimed at mitigating  thermal decoherence pathways such as those involving vibronic coupling in the next-generation of molecular spin qubits for quantum technologies.
Both the site-specific spectroscopic technique and the first-principles theoretical approach demonstrated herein can be generally applied to identify crystal field excitations in other $f$-block complexes, regardless of the coordination environment. This approach %opens the door to
paves the way toward
a high temperature frontier for molecular nanomagnets where first-principles prediction and  experimental optimization  %the high-temperature frontier of
%molecular nanomagnets and the
advance the development of molecular spin qubits with improved lifetimes and high operating temperatures.

%%%%%%%%%%%%%%%%%%%%%%%%%%%%%%%%%%%%
\section*{Methods}
%%%%%%%%%%%%%%%%%%%%%%%%%%%%%%%%%%%%

\paragraph{Crystal growth and sample preparation:} Single crystals of Na$_{9}$[Ho(W$_5$O$_{18}$)$_2$]$\cdot$35H$_2$O were grown by solution techniques. \cite{Shiddiq2016} The non-magnetic Y$^{3+}$ analog was prepared as well.  In order to control the optical density, we combined the crystals with a paraffin matrix which is transparent in the far infrared region.

\paragraph{Far infrared spectroscopy:} We employed a Bruker 113v Fourier transform infrared spectrometer equipped with a bolometer detector to reveal the far infrared response of \HoWEn. The measured transmittance $T$($\omega$) was converted to absorption as
$\alpha(\omega) = -\frac{1}{\textit{hd}}\ln \left(T(\omega\right)$.
 Here, $\textit{h}$ is the loading,  and  $\textit{d}$ is the thickness.
An open flow cryostat was used for temperature control.
 Magneto-infrared work was performed at the National High Magnetic Field Laboratory in Tallahassee, FL using a Bruker 66V/S spectrometer and a 35 T resistive magnet. Absorption differences are ideal for revealing small field-induced spectral changes. They are  calculated as: $\Delta\alpha({\omega},H) = \alpha({\omega},H) - \alpha({\omega},H$ = 0~T).  Here, $H$ is applied magnetic field, and $\omega$ is the frequency.

\paragraph{DFT calculations:} The structural optimization of the crystallographic coordinates (in vacuum) and the vibrational modes calculations were carried out at DFT level using the Gaussian16 package in its revision A.03. \cite{Gaussian} Vibrational frequency calculations were performed using both the fully optimized structure and the X-ray crystal structure with no optimization. The PBE0 hybrid exchange-correlation functional was used for both optimization and frequency calculations in combination with Stuttgart RSC ANO basis set with effective core potential (ECP) for the Ho$^{3+}$ cation. CRENBL basis set have been used for W with corresponding ECP potential and 6-31G(d,p) basis set had been used for oxygen. An `ultra-fine' integration grid and `very tight' SCF convergence criterion were applied. Dispersion effects were taken into account using the empirical GD3BJ dispersion correction.

\paragraph{\emph{Ab initio} calculations:}The time-independent electronic structure was computed using the multireference Complete Active Self-Consistent Field Spin-Orbit (CASSCF-SO) method as implemented in the OpenMOLCAS program package (version 18.09).\cite{fdez2019openmolcas} The molecular geometry was extracted from the crystal structure determined by Single Crystal X-ray diffraction and was fully optimized at density functional theory (DFT) level ({\it vide infra}). In addition, the electronic structure of the molecular geometry with no prior optimization was calculated. Scalar relativistic effects were taken into account with the Douglas–Kroll–Hess
transformation using the relativistically contracted atomic natural orbital ANO-RCC basis set with VDZP quality for all atoms. The active space
consisted of 10 electrons on the 7 f-orbitals of Ho$^{3+}$ ion. The
molecular orbitals were optimized at the CASSCF level in a state-average (SA)
over 35 quintets of the ground state term ($L$ = 6
for Ho$^{3+}$). The wave functions obtained at CASSCF were then mixed by
spin-orbit coupling by means of the RASSI approach.
The combined effect of the crystal field and the spin orbit coupling were computed using SINGLE-ANISO module.\cite{ungur2017ab}

\paragraph{Vibronic coupling:} For vibronic-coupling-dependent IR absorption spectra, we defined an effective Hamiltonian described as:
\begin{equation}\label{effec}
	\hat{H}_{eff} = \hat{H}_{S} + \hat{H}_{vib} + \hat{H}_{S-vib}
\end{equation}
where $\hat{H}_{S}$,  $\hat{H}_{vib}$ and $\hat{H}_{S-vib}$ correspond to the system, the vibration (bath) and the spin-vibration
Hamiltonian, which can be written as:
\begin{equation}
	\hat{H}_{S} = \sum_{k=2,4,6}\sum_{q=-k}^{k} B_k^q \hat{O}_k^q (J)
\end{equation}
\begin{equation}
	\hat{H}_{vib} = \sum_{j} \hbar\omega_j(n_j+1/2)
\end{equation}
\begin{equation}
	\hat{H}_{S-vib} = \sum_{j}\sum_{kq}\hat{Q}_j\left(\frac{\partial B_k^q}{\partial Q_j}\right)_0\hat{O}_k^q (J)
\end{equation}
where $B_k^q$ and $\hat{O}_k^q (J)$ correspond to the crystal field parameter and the Stevens operator, respectively. $j$ is an index running over the vibrations and $n_j$ correspond to the vibrational level of mode $j$. $\hat{Q}_j$ denotes the vibrational coordinate and the term $\left(\frac{\partial B_k^q}{\partial Q_j}\right)_0$ is the vibronic coupling for a given vibrational mode $j$. To compute this term, we carried out {\it ab initio} calculations at CASSCF-SO level.
The introduction of vibronic coupling is key to compute the IR absortion spectra, since due to the mixing more transitions will be optically allowed. Complete details can be found as Supplementary Information.

%%%%%%%%%%%%%%%%%%%%%%%%%%%%%%%%
\section*{Acknowledgements}
%%%%%%%%%%%%%%%%%%%%%%%%%%%%%%%%

Research at the University of Tennessee is supported by the National Science Foundation
(DMR-1707846) and the Materials Research Fund through the University of Tennessee. Research at Universitat de Valencia is supported by the EU (ERC-2014-CoG-647301 DECRESIM,
ERC-2018-AdG-788222 MOL-2D, the QUANTERA
project SUMO, and FET-OPEN grant 862893 FATMOLS); the Spanish MCIU (grant CTQ2017-89993 and PGC2018-099568-B-I00 co-financed by FEDER, grant MAT2017-89528; the Unit of excellence `María de Maeztu'
CEX2019-000919-M); the Generalitat Valenciana (Prometeo Program of Excellence, SEJI/2018/035 and grant CDEIGENT/2019/022).
A portion of this work was performed at the National High Magnetic Field Laboratory, which is funded by National Science Foundation Cooperative Agreement No. DMR-1644779 and the State of Florida.

\bibliography{achemso}

\providecommand{\latin}[1]{#1}
\makeatletter
\providecommand{\doi}
  {\begingroup\let\do\@makeother\dospecials
  \catcode`\{=1 \catcode`\}=2 \doi@aux}
\providecommand{\doi@aux}[1]{\endgroup\texttt{#1}}
\makeatother
\providecommand*\mcitethebibliography{\thebibliography}
\csname @ifundefined\endcsname{endmcitethebibliography}
  {\let\endmcitethebibliography\endthebibliography}{}
\begin{mcitethebibliography}{47}
\providecommand*\natexlab[1]{#1}
\providecommand*\mciteSetBstSublistMode[1]{}
\providecommand*\mciteSetBstMaxWidthForm[2]{}
\providecommand*\mciteBstWouldAddEndPuncttrue
  {\def\EndOfBibitem{\unskip.}}
\providecommand*\mciteBstWouldAddEndPunctfalse
  {\let\EndOfBibitem\relax}
\providecommand*\mciteSetBstMidEndSepPunct[3]{}
\providecommand*\mciteSetBstSublistLabelBeginEnd[3]{}
\providecommand*\EndOfBibitem{}
\mciteSetBstSublistMode{f}
\mciteSetBstMaxWidthForm{subitem}{(\alph{mcitesubitemcount})}
\mciteSetBstSublistLabelBeginEnd
  {\mcitemaxwidthsubitemform\space}
  {\relax}
  {\relax}

\bibitem[Coronado(2020)]{coronado2020molecular}
Coronado,~E. Molecular magnetism: from chemical design to spin control in
  molecules, materials and devices. \emph{Nature Reviews Materials}
  \textbf{2020}, \emph{5}, 87--104\relax
\mciteBstWouldAddEndPuncttrue
\mciteSetBstMidEndSepPunct{\mcitedefaultmidpunct}
{\mcitedefaultendpunct}{\mcitedefaultseppunct}\relax
\EndOfBibitem
\bibitem[Gaita-Ari{\~n}o \latin{et~al.}(2019)Gaita-Ari{\~n}o, Luis, Hill, and
  Coronado]{gaita2019molecular}
Gaita-Ari{\~n}o,~A.; Luis,~F.; Hill,~S.; Coronado,~E. Molecular spins for
  quantum computation. \emph{Nature Chemistry} \textbf{2019}, \emph{11},
  301--309\relax
\mciteBstWouldAddEndPuncttrue
\mciteSetBstMidEndSepPunct{\mcitedefaultmidpunct}
{\mcitedefaultendpunct}{\mcitedefaultseppunct}\relax
\EndOfBibitem
\bibitem[Atzori and Sessoli(2019)Atzori, and Sessoli]{atzori2019second}
Atzori,~M.; Sessoli,~R. The second quantum revolution: role and challenges of
  molecular chemistry. \emph{Journal of the American Chemical Society}
  \textbf{2019}, \emph{141}, 11339--11352\relax
\mciteBstWouldAddEndPuncttrue
\mciteSetBstMidEndSepPunct{\mcitedefaultmidpunct}
{\mcitedefaultendpunct}{\mcitedefaultseppunct}\relax
\EndOfBibitem
\bibitem[Woodruff \latin{et~al.}(2013)Woodruff, Winpenny, and
  Layfield]{Woodruff2013}
Woodruff,~D.~N.; Winpenny,~R.~E.; Layfield,~R.~A. Lanthanide single-molecule
  magnets. \emph{Chemical Reviews} \textbf{2013}, \emph{113}, 5110--5148\relax
\mciteBstWouldAddEndPuncttrue
\mciteSetBstMidEndSepPunct{\mcitedefaultmidpunct}
{\mcitedefaultendpunct}{\mcitedefaultseppunct}\relax
\EndOfBibitem
\bibitem[Guo \latin{et~al.}(2018)Guo, Day, Chen, Tong, Mansikkam{\"a}ki, and
  Layfield]{Guo2018}
Guo,~F.-S.; Day,~B.~M.; Chen,~Y.-C.; Tong,~M.-L.; Mansikkam{\"a}ki,~A.;
  Layfield,~R.~A. Magnetic hysteresis up to 80 kelvin in a dysprosium
  metallocene single-molecule magnet. \emph{Science} \textbf{2018}, \emph{362},
  1400--1403\relax
\mciteBstWouldAddEndPuncttrue
\mciteSetBstMidEndSepPunct{\mcitedefaultmidpunct}
{\mcitedefaultendpunct}{\mcitedefaultseppunct}\relax
\EndOfBibitem
\bibitem[Zadrozny \latin{et~al.}(2015)Zadrozny, Niklas, Poluektov, and
  Freedman]{zadrozny2015millisecond}
Zadrozny,~J.~M.; Niklas,~J.; Poluektov,~O.~G.; Freedman,~D.~E. Millisecond
  coherence time in a tunable molecular electronic spin qubit. \emph{ACS
  Central Science} \textbf{2015}, \emph{1}, 488--492\relax
\mciteBstWouldAddEndPuncttrue
\mciteSetBstMidEndSepPunct{\mcitedefaultmidpunct}
{\mcitedefaultendpunct}{\mcitedefaultseppunct}\relax
\EndOfBibitem
\bibitem[Atzori \latin{et~al.}(2016)Atzori, Tesi, Morra, Chiesa, Sorace, and
  Sessoli]{atzori2016room}
Atzori,~M.; Tesi,~L.; Morra,~E.; Chiesa,~M.; Sorace,~L.; Sessoli,~R.
  Room-temperature quantum coherence and rabi oscillations in vanadyl
  phthalocyanine: toward multifunctional molecular spin qubits. \emph{Journal
  of the American Chemical Society} \textbf{2016}, \emph{138}, 2154--2157\relax
\mciteBstWouldAddEndPuncttrue
\mciteSetBstMidEndSepPunct{\mcitedefaultmidpunct}
{\mcitedefaultendpunct}{\mcitedefaultseppunct}\relax
\EndOfBibitem
\bibitem[Bader \latin{et~al.}(2014)Bader, Dengler, Lenz, Endeward, Jiang,
  Neugebauer, and Van~Slageren]{bader2014room}
Bader,~K.; Dengler,~D.; Lenz,~S.; Endeward,~B.; Jiang,~S.-D.; Neugebauer,~P.;
  Van~Slageren,~J. Room temperature quantum coherence in a potential molecular
  qubit. \emph{Nature Communications} \textbf{2014}, \emph{5}, 1--5\relax
\mciteBstWouldAddEndPuncttrue
\mciteSetBstMidEndSepPunct{\mcitedefaultmidpunct}
{\mcitedefaultendpunct}{\mcitedefaultseppunct}\relax
\EndOfBibitem
\bibitem[Shiddiq \latin{et~al.}(2016)Shiddiq, Komijani, Duan, Gaita-Ari{\~n}o,
  Coronado, and Hill]{Shiddiq2016}
Shiddiq,~M.; Komijani,~D.; Duan,~Y.; Gaita-Ari{\~n}o,~A.; Coronado,~E.;
  Hill,~S. Enhancing coherence in molecular spin qubits via atomic clock
  transitions. \emph{Nature} \textbf{2016}, \emph{531}, 348--351\relax
\mciteBstWouldAddEndPuncttrue
\mciteSetBstMidEndSepPunct{\mcitedefaultmidpunct}
{\mcitedefaultendpunct}{\mcitedefaultseppunct}\relax
\EndOfBibitem
\bibitem[Hackerm{\"u}ller \latin{et~al.}(2004)Hackerm{\"u}ller, Hornberger,
  Brezger, Zeilinger, and Arndt]{Hackermuller2004}
Hackerm{\"u}ller,~L.; Hornberger,~K.; Brezger,~B.; Zeilinger,~A.; Arndt,~M.
  Decoherence of matter waves by thermal emission of radiation. \emph{Nature}
  \textbf{2004}, \emph{427}, 711--714\relax
\mciteBstWouldAddEndPuncttrue
\mciteSetBstMidEndSepPunct{\mcitedefaultmidpunct}
{\mcitedefaultendpunct}{\mcitedefaultseppunct}\relax
\EndOfBibitem
\bibitem[Chirolli and Burkard(2008)Chirolli, and Burkard]{Chirolli2008}
Chirolli,~L.; Burkard,~G. Decoherence in solid-state qubits. \emph{Advances in
  Physics} \textbf{2008}, \emph{57}, 225--285\relax
\mciteBstWouldAddEndPuncttrue
\mciteSetBstMidEndSepPunct{\mcitedefaultmidpunct}
{\mcitedefaultendpunct}{\mcitedefaultseppunct}\relax
\EndOfBibitem
\bibitem[Graham \latin{et~al.}(2014)Graham, Zadrozny, Shiddiq, Anderson,
  Fataftah, Hill, and Freedman]{Graham2014}
Graham,~M.~J.; Zadrozny,~J.~M.; Shiddiq,~M.; Anderson,~J.~S.; Fataftah,~M.~S.;
  Hill,~S.; Freedman,~D.~E. Influence of electronic spin and spin--orbit
  coupling on decoherence in mononuclear transition metal complexes.
  \emph{Journal of the American Chemical Society} \textbf{2014}, \emph{136},
  7623--7626\relax
\mciteBstWouldAddEndPuncttrue
\mciteSetBstMidEndSepPunct{\mcitedefaultmidpunct}
{\mcitedefaultendpunct}{\mcitedefaultseppunct}\relax
\EndOfBibitem
\bibitem[Goodwin \latin{et~al.}(2017)Goodwin, Reta, Ortu, Chilton, and
  Mills]{Goodwin2017}
Goodwin,~C.~A.; Reta,~D.; Ortu,~F.; Chilton,~N.~F.; Mills,~D.~P. Synthesis and
  electronic structures of heavy lanthanide metallocenium cations.
  \emph{Journal of the American Chemical Society} \textbf{2017}, \emph{139},
  18714--18724\relax
\mciteBstWouldAddEndPuncttrue
\mciteSetBstMidEndSepPunct{\mcitedefaultmidpunct}
{\mcitedefaultendpunct}{\mcitedefaultseppunct}\relax
\EndOfBibitem
\bibitem[Shiozaki \latin{et~al.}(1996)Shiozaki, Inagaki, Nishino, Nishio,
  Maekawa, Kominami, and Kera]{Shiozaki1996}
Shiozaki,~R.; Inagaki,~A.; Nishino,~A.; Nishio,~E.; Maekawa,~M.; Kominami,~H.;
  Kera,~Y. Spectroscopic investigation of a series of sodium lanthanide
  decatungstates, Na$_7$H$_2$Ln(III)(W$_5$O$_{18}$)$_2${\textperiodcentered}
  $n$H$_2$O (Ln: La-Yb): the contribution of 4f$^{n}$ electrons to bonding
  interaction among Ln(III) and polyoxotungstates. \emph{Journal of Alloys and
  Compounds} \textbf{1996}, \emph{234}, 193--198\relax
\mciteBstWouldAddEndPuncttrue
\mciteSetBstMidEndSepPunct{\mcitedefaultmidpunct}
{\mcitedefaultendpunct}{\mcitedefaultseppunct}\relax
\EndOfBibitem
\bibitem[AlDamen \latin{et~al.}(2009)AlDamen, Cardona-Serra, Clemente-Juan,
  Coronado, Gaita-Ari{\~n}o, Mart{\'\i}-Gastaldo, Luis, and
  Montero]{AlDamen2009}
AlDamen,~M.~A.; Cardona-Serra,~S.; Clemente-Juan,~J.~M.; Coronado,~E.;
  Gaita-Ari{\~n}o,~A.; Mart{\'\i}-Gastaldo,~C.; Luis,~F.; Montero,~O.
  Mononuclear lanthanide single molecule magnets based on the polyoxometalates
  [Ln(W$_5$O$_{18}$)$_2$]$^{9-}$ and
  [Ln($\beta_2$-SiW$_{11}$O$_{39}$)$_2$]$^{13-}$ (Ln$^{III}$ = Tb, Dy, Ho, Er,
  Tm, and Yb). \emph{Inorganic Chemistry} \textbf{2009}, \emph{48},
  3467--3479\relax
\mciteBstWouldAddEndPuncttrue
\mciteSetBstMidEndSepPunct{\mcitedefaultmidpunct}
{\mcitedefaultendpunct}{\mcitedefaultseppunct}\relax
\EndOfBibitem
\bibitem[Gu \latin{et~al.}(2018)Gu, Chen, Wu, Zheng, and Li]{Gu2018}
Gu,~Y.-N.; Chen,~Y.; Wu,~Y.-L.; Zheng,~S.-T.; Li,~X.-X. A series of
  banana-shaped 3d-4f heterometallic cluster substituted polyoxometalates:
  syntheses, crystal structures, and magnetic properties. \emph{Inorganic
  Chemistry} \textbf{2018}, \emph{57}, 2472--2479\relax
\mciteBstWouldAddEndPuncttrue
\mciteSetBstMidEndSepPunct{\mcitedefaultmidpunct}
{\mcitedefaultendpunct}{\mcitedefaultseppunct}\relax
\EndOfBibitem
\bibitem[AlDamen \latin{et~al.}(2008)AlDamen, Clemente-Juan, Coronado,
  Mart{\'\i}-Gastaldo, and Gaita-Ari{\~n}o]{AlDamen2008}
AlDamen,~M.~A.; Clemente-Juan,~J.~M.; Coronado,~E.; Mart{\'\i}-Gastaldo,~C.;
  Gaita-Ari{\~n}o,~A. Mononuclear lanthanide single-molecule magnets based on
  polyoxometalates. \emph{Journal of the American Chemical Society}
  \textbf{2008}, \emph{130}, 8874--8875\relax
\mciteBstWouldAddEndPuncttrue
\mciteSetBstMidEndSepPunct{\mcitedefaultmidpunct}
{\mcitedefaultendpunct}{\mcitedefaultseppunct}\relax
\EndOfBibitem
\bibitem[Li \latin{et~al.}(2006)Li, Parkin, Wang, Yee, Cl{\'e}rac, Wernsdorfer,
  and Holmes]{Li2006}
Li,~D.; Parkin,~S.; Wang,~G.; Yee,~G.~T.; Cl{\'e}rac,~R.; Wernsdorfer,~W.;
  Holmes,~S.~M. An $S$ = 6 cyanide-bridged octanuclear
  Fe$^{III}$$_4$Ni$^{II}$$_4$ complex that exhibits slow relaxation of the
  magnetization. \emph{Journal of the American Chemical Society} \textbf{2006},
  \emph{128}, 4214--4215\relax
\mciteBstWouldAddEndPuncttrue
\mciteSetBstMidEndSepPunct{\mcitedefaultmidpunct}
{\mcitedefaultendpunct}{\mcitedefaultseppunct}\relax
\EndOfBibitem
\bibitem[Ghosh \latin{et~al.}(2012)Ghosh, Datta, Friend, Cardona-Serra,
  Gaita-Ari{\~n}o, Coronado, and Hill]{Ghosh2012}
Ghosh,~S.; Datta,~S.; Friend,~L.; Cardona-Serra,~S.; Gaita-Ari{\~n}o,~A.;
  Coronado,~E.; Hill,~S. Multi-frequency EPR studies of a mononuclear holmium
  single-molecule magnet based on the polyoxometalate
  [Ho$^{III}$(W$_5$O$_{18}$)$_2$]$^{9-}$. \emph{Dalton Transactions}
  \textbf{2012}, \emph{41}, 13697--13704\relax
\mciteBstWouldAddEndPuncttrue
\mciteSetBstMidEndSepPunct{\mcitedefaultmidpunct}
{\mcitedefaultendpunct}{\mcitedefaultseppunct}\relax
\EndOfBibitem
\bibitem[Liu \latin{et~al.}(2020)Liu, Mrozek, Duan, Ullah, Baldov{\'\i},
  Coronado, Gaita-Ari{\~n}o, and Ardavan]{liu2020quantum}
Liu,~J.; Mrozek,~J.; Duan,~Y.; Ullah,~A.; Baldov{\'\i},~J.~J.; Coronado,~E.;
  Gaita-Ari{\~n}o,~A.; Ardavan,~A. Quantum coherent spin-electric control in
  molecular nanomagnets. \emph{arXiv preprint arXiv:2005.01029} \textbf{2020},
  \relax
\mciteBstWouldAddEndPunctfalse
\mciteSetBstMidEndSepPunct{\mcitedefaultmidpunct}
{}{\mcitedefaultseppunct}\relax
\EndOfBibitem
\bibitem[Vonci \latin{et~al.}(2017)Vonci, Giansiracusa, Van~den Heuvel, Gable,
  Moubaraki, Murray, Yu, Mole, Soncini, and Boskovic]{Vonci2017}
Vonci,~M.; Giansiracusa,~M.~J.; Van~den Heuvel,~W.; Gable,~R.~W.;
  Moubaraki,~B.; Murray,~K.~S.; Yu,~D.; Mole,~R.~A.; Soncini,~A.; Boskovic,~C.
  Magnetic excitations in polyoxotungstate-supported lanthanoid single-molecule
  magnets: an inelastic neutron scattering and ab initio study. \emph{Inorganic
  Chemistry} \textbf{2017}, \emph{56}, 378--394\relax
\mciteBstWouldAddEndPuncttrue
\mciteSetBstMidEndSepPunct{\mcitedefaultmidpunct}
{\mcitedefaultendpunct}{\mcitedefaultseppunct}\relax
\EndOfBibitem
\bibitem[Baldov{\'\i} \latin{et~al.}(2014)Baldov{\'\i}, Clemente-Juan,
  Coronado, Duan, Gaita-Ari\~no, and Gim\'enez-Saiz]{Baldovi2014}
Baldov{\'\i},~J.~J.; Clemente-Juan,~J.~M.; Coronado,~E.; Duan,~Y.;
  Gaita-Ari\~no,~A.; Gim\'enez-Saiz, Construction of a general library for the
  rational design of nanomagnets and spin qubits based on mononuclear f-block
  complexes. The polyoxometalate case. \emph{Inorganic Chemistry}
  \textbf{2014}, \emph{53}, 9976--9980\relax
\mciteBstWouldAddEndPuncttrue
\mciteSetBstMidEndSepPunct{\mcitedefaultmidpunct}
{\mcitedefaultendpunct}{\mcitedefaultseppunct}\relax
\EndOfBibitem
\bibitem[Brinzari \latin{et~al.}(2013)Brinzari, Chen, Sun, Liu, Tung, Wang,
  Schlueter, Singleton, Manson, Whangbo, \latin{et~al.} others]{Brinzari2013}
Brinzari,~T.; Chen,~P.; Sun,~Q.-C.; Liu,~J.; Tung,~L.-C.; Wang,~Y.;
  Schlueter,~J.; Singleton,~J.; Manson,~J.~L.; Whangbo,~M.-H., \latin{et~al.}
  Quantum critical transition amplifies magnetoelastic coupling in
  Mn[N(CN)$_2$]$_2$. \emph{Physical Review Letters} \textbf{2013}, \emph{110},
  237202\relax
\mciteBstWouldAddEndPuncttrue
\mciteSetBstMidEndSepPunct{\mcitedefaultmidpunct}
{\mcitedefaultendpunct}{\mcitedefaultseppunct}\relax
\EndOfBibitem
\bibitem[Brinzari \latin{et~al.}(2013)Brinzari, Haraldsen, Chen, Sun, Kim,
  Tung, Litvinchuk, Schlueter, Smirnov, Manson, \latin{et~al.}
  others]{Brinzari2013b}
Brinzari,~T.~V.; Haraldsen,~J.~T.; Chen,~P.; Sun,~Q.-C.; Kim,~Y.; Tung,~L.-C.;
  Litvinchuk,~A.~P.; Schlueter,~J.~A.; Smirnov,~D.; Manson,~J.~L.,
  \latin{et~al.}  Electron-phonon and magnetoelastic interactions in
  ferromagnetic Co[N(CN)$_2$]$_2$. \emph{Physical Review Letters}
  \textbf{2013}, \emph{111}, 047202\relax
\mciteBstWouldAddEndPuncttrue
\mciteSetBstMidEndSepPunct{\mcitedefaultmidpunct}
{\mcitedefaultendpunct}{\mcitedefaultseppunct}\relax
\EndOfBibitem
\bibitem[H{\"a}rtle \latin{et~al.}(2011)H{\"a}rtle, Butzin, Rubio-Pons, and
  Thoss]{Hartle2011}
H{\"a}rtle,~R.; Butzin,~M.; Rubio-Pons,~O.; Thoss,~M. Quantum interference and
  decoherence in single-molecule junctions: how vibrations induce electrical
  current. \emph{Physical Review Letters} \textbf{2011}, \emph{107},
  046802\relax
\mciteBstWouldAddEndPuncttrue
\mciteSetBstMidEndSepPunct{\mcitedefaultmidpunct}
{\mcitedefaultendpunct}{\mcitedefaultseppunct}\relax
\EndOfBibitem
\bibitem[Ballmann \latin{et~al.}(2012)Ballmann, H{\"a}rtle, Coto, Elbing,
  Mayor, Bryce, Thoss, and Weber]{Ballmann2012}
Ballmann,~S.; H{\"a}rtle,~R.; Coto,~P.~B.; Elbing,~M.; Mayor,~M.; Bryce,~M.~R.;
  Thoss,~M.; Weber,~H.~B. Experimental evidence for quantum interference and
  vibrationally induced decoherence in single-molecule junctions.
  \emph{Physical Review Letters} \textbf{2012}, \emph{109}, 056801\relax
\mciteBstWouldAddEndPuncttrue
\mciteSetBstMidEndSepPunct{\mcitedefaultmidpunct}
{\mcitedefaultendpunct}{\mcitedefaultseppunct}\relax
\EndOfBibitem
\bibitem[Escalera-Moreno \latin{et~al.}(2018)Escalera-Moreno, Baldov{\'\i},
  Gaita-Ari{\~n}o, and Coronado]{escalera2018spin}
Escalera-Moreno,~L.; Baldov{\'\i},~J.~J.; Gaita-Ari{\~n}o,~A.; Coronado,~E.
  Spin states, vibrations and spin relaxation in molecular nanomagnets and spin
  qubits: a critical perspective. \emph{Chemical Science} \textbf{2018},
  \emph{9}, 3265--3275\relax
\mciteBstWouldAddEndPuncttrue
\mciteSetBstMidEndSepPunct{\mcitedefaultmidpunct}
{\mcitedefaultendpunct}{\mcitedefaultseppunct}\relax
\EndOfBibitem
\bibitem[Ullah \latin{et~al.}(2019)Ullah, Cerd\'a, Baldov{\'\i}, Varganov,
  Arag\'o, and Gaita-Ari{\~n}o]{ullah2019silico}
Ullah,~A.; Cerd\'a,~J.; Baldov{\'\i},~J.~J.; Varganov,~S.~A.; Arag\'o,~J.;
  Gaita-Ari{\~n}o,~A. In silico molecular engineering of dysprosocenium-based
  complexes to decouple spin energy levels from molecular vibrations. \emph{The
  Journal of Physical Chemistry Letters} \textbf{2019}, \emph{10},
  7678--7683\relax
\mciteBstWouldAddEndPuncttrue
\mciteSetBstMidEndSepPunct{\mcitedefaultmidpunct}
{\mcitedefaultendpunct}{\mcitedefaultseppunct}\relax
\EndOfBibitem
\bibitem[Garlatti \latin{et~al.}(2020)Garlatti, Tesi, Lunghi, Atzori, Voneshen,
  Santini, Sanvito, Guidi, Sessoli, and Carretta]{garlatti2020unveiling}
Garlatti,~E.; Tesi,~L.; Lunghi,~A.; Atzori,~M.; Voneshen,~D.; Santini,~P.;
  Sanvito,~S.; Guidi,~T.; Sessoli,~R.; Carretta,~S. Unveiling phonons in a
  molecular qubit with four-dimensional inelastic neutron scattering and
  density functional theory. \emph{Nature Communications} \textbf{2020},
  \emph{11}, 1--10\relax
\mciteBstWouldAddEndPuncttrue
\mciteSetBstMidEndSepPunct{\mcitedefaultmidpunct}
{\mcitedefaultendpunct}{\mcitedefaultseppunct}\relax
\EndOfBibitem
\bibitem[Yu \latin{et~al.}(2020)Yu, Von~Kugelgen, Krzyaniak, Ji, Dichtel,
  Wasielewski, and Freedman]{yu2020spin}
Yu,~C.-J.; Von~Kugelgen,~S.; Krzyaniak,~M.~D.; Ji,~W.; Dichtel,~W.~R.;
  Wasielewski,~M.~R.; Freedman,~D.~E. Spin and Phonon Design in Modular Arrays
  of Molecular Qubits. \emph{Chemistry of Materials} \textbf{2020}, \emph{32},
  10200--10206\relax
\mciteBstWouldAddEndPuncttrue
\mciteSetBstMidEndSepPunct{\mcitedefaultmidpunct}
{\mcitedefaultendpunct}{\mcitedefaultseppunct}\relax
\EndOfBibitem
\bibitem[Marx \latin{et~al.}(2014)Marx, Moro, D{\"o}rfel, Ungur, Waters, Jiang,
  Orlita, Taylor, Frey, Chibotaru, \latin{et~al.} others]{Marx2014}
Marx,~R.; Moro,~F.; D{\"o}rfel,~M.; Ungur,~L.; Waters,~M.; Jiang,~S.-D.;
  Orlita,~M.; Taylor,~J.; Frey,~W.; Chibotaru,~L., \latin{et~al.}
  Spectroscopic determination of crystal field splittings in lanthanide double
  deckers. \emph{Chemical Science} \textbf{2014}, \emph{5}, 3287--3293\relax
\mciteBstWouldAddEndPuncttrue
\mciteSetBstMidEndSepPunct{\mcitedefaultmidpunct}
{\mcitedefaultendpunct}{\mcitedefaultseppunct}\relax
\EndOfBibitem
\bibitem[Kumar \latin{et~al.}(2016)Kumar, Xiao, Nair, Voigt, Schmitz,
  Chatterji, Jalarvo, and Br{\"u}ckel]{Kumar2016}
Kumar,~C.; Xiao,~Y.; Nair,~H.; Voigt,~J.; Schmitz,~B.; Chatterji,~T.;
  Jalarvo,~N.; Br{\"u}ckel,~T. Hyperfine and crystal field interactions in
  multiferroic HoCrO$_3$. \emph{Journal of Physics: Condensed Matter}
  \textbf{2016}, \emph{28}, 476001\relax
\mciteBstWouldAddEndPuncttrue
\mciteSetBstMidEndSepPunct{\mcitedefaultmidpunct}
{\mcitedefaultendpunct}{\mcitedefaultseppunct}\relax
\EndOfBibitem
\bibitem[Ungur and Chibotaru(2017)Ungur, and Chibotaru]{Ungur2017}
Ungur,~L.; Chibotaru,~L.~F. Ab initio crystal field for lanthanides.
  \emph{Chemistry--A European Journal} \textbf{2017}, \emph{23},
  3708--3718\relax
\mciteBstWouldAddEndPuncttrue
\mciteSetBstMidEndSepPunct{\mcitedefaultmidpunct}
{\mcitedefaultendpunct}{\mcitedefaultseppunct}\relax
\EndOfBibitem
\bibitem[Newman(1971)]{Newman1971}
Newman,~D.~J. Theory of lanthanide crystal fields. \emph{Advances in Physics}
  \textbf{1971}, \emph{20}, 197--256\relax
\mciteBstWouldAddEndPuncttrue
\mciteSetBstMidEndSepPunct{\mcitedefaultmidpunct}
{\mcitedefaultendpunct}{\mcitedefaultseppunct}\relax
\EndOfBibitem
\bibitem[Hallmen \latin{et~al.}(2018)Hallmen, Rauhut, Stoll, Mitrushchenkov,
  and van Slageren]{Hallmen2018}
Hallmen,~P.~P.; Rauhut,~G.; Stoll,~H.; Mitrushchenkov,~A.; van Slageren,~J.
  Crystal field splittings in lanthanide complexes: inclusion of correlation
  effects beyond second order perturbation theory. \emph{Journal of Chemical
  Theory and Computation} \textbf{2018}, \emph{14}, 3998--4009\relax
\mciteBstWouldAddEndPuncttrue
\mciteSetBstMidEndSepPunct{\mcitedefaultmidpunct}
{\mcitedefaultendpunct}{\mcitedefaultseppunct}\relax
\EndOfBibitem
\bibitem[Jones \latin{et~al.}(2001)Jones, Varughese, Olejniczak, Pigos,
  Musfeldt, Landee, Turnbull, and Carr]{Jones2001}
Jones,~B.; Varughese,~P.; Olejniczak,~I.; Pigos,~J.; Musfeldt,~J.; Landee,~C.;
  Turnbull,~M.; Carr,~G. Vibrational properties of the one-dimensional, S=1/2,
  Heisenberg antiferromagnet copper pyrazine dinitrate. \emph{Chemistry of
  Materials} \textbf{2001}, \emph{13}, 2127--2134\relax
\mciteBstWouldAddEndPuncttrue
\mciteSetBstMidEndSepPunct{\mcitedefaultmidpunct}
{\mcitedefaultendpunct}{\mcitedefaultseppunct}\relax
\EndOfBibitem
\bibitem[Casto \latin{et~al.}(2015)Casto, Clune, Yokosuk, Musfeldt, Williams,
  Zhuang, Lin, Xiao, Hennig, Sales, \latin{et~al.} others]{Casto2015}
Casto,~L.; Clune,~A.; Yokosuk,~M.; Musfeldt,~J.; Williams,~T.; Zhuang,~H.;
  Lin,~M.-W.; Xiao,~K.; Hennig,~R.; Sales,~B., \latin{et~al.}  Strong
  spin-lattice coupling in CrSiTe$_3$. \emph{APL Materials} \textbf{2015},
  \emph{3}, 041515\relax
\mciteBstWouldAddEndPuncttrue
\mciteSetBstMidEndSepPunct{\mcitedefaultmidpunct}
{\mcitedefaultendpunct}{\mcitedefaultseppunct}\relax
\EndOfBibitem
\bibitem[Sethi \latin{et~al.}(2019)Sethi, Slimak, Kolodiazhnyi, and
  Cooper]{Sethi2019}
Sethi,~A.; Slimak,~J.; Kolodiazhnyi,~T.; Cooper,~S. Emergent vibronic
  excitations in the magnetodielectric regime of Ce$_2$O$_3$. \emph{Physical
  Review Letters} \textbf{2019}, \emph{122}, 177601\relax
\mciteBstWouldAddEndPuncttrue
\mciteSetBstMidEndSepPunct{\mcitedefaultmidpunct}
{\mcitedefaultendpunct}{\mcitedefaultseppunct}\relax
\EndOfBibitem
\bibitem[Tang \latin{et~al.}(2020)Tang, Su, Lao, Bao, Zhu, and Xu]{Tang2020}
Tang,~F.; Su,~Z.; Lao,~X.; Bao,~Y.; Zhu,~D.; Xu,~S. The key roles of 4f-level
  splitting and vibronic coupling in the high-efficiency luminescence of
  Ce$^{3+}$ ion in LuAG transparent ceramic phosphors. \emph{Journal of
  Luminescence} \textbf{2020}, 117360\relax
\mciteBstWouldAddEndPuncttrue
\mciteSetBstMidEndSepPunct{\mcitedefaultmidpunct}
{\mcitedefaultendpunct}{\mcitedefaultseppunct}\relax
\EndOfBibitem
\bibitem[Tayal \latin{et~al.}(1980)Tayal, Srivastava, Khandelwal, and
  Bist]{Tayal1980}
Tayal,~V.; Srivastava,~B.; Khandelwal,~D.; Bist,~H. Librational modes of
  crystal water in hydrated solids. \emph{Applied Spectroscopy Reviews}
  \textbf{1980}, \emph{16}, 43--134\relax
\mciteBstWouldAddEndPuncttrue
\mciteSetBstMidEndSepPunct{\mcitedefaultmidpunct}
{\mcitedefaultendpunct}{\mcitedefaultseppunct}\relax
\EndOfBibitem
\bibitem[Dominguez-Vidal \latin{et~al.}(2007)Dominguez-Vidal, Kaun,
  Ayora-Ca{\~n}ada, and Lendl]{Dominguez2007}
Dominguez-Vidal,~A.; Kaun,~N.; Ayora-Ca{\~n}ada,~M.~J.; Lendl,~B. Probing
  intermolecular interactions in water/ionic liquid mixtures by far-infrared
  spectroscopy. \emph{The Journal of Physical Chemistry B} \textbf{2007},
  \emph{111}, 4446--4452\relax
\mciteBstWouldAddEndPuncttrue
\mciteSetBstMidEndSepPunct{\mcitedefaultmidpunct}
{\mcitedefaultendpunct}{\mcitedefaultseppunct}\relax
\EndOfBibitem
\bibitem[Ishikawa \latin{et~al.}(2003)Ishikawa, Sugita, Ishikawa, Koshihara,
  and Kaizu]{ishikawa2003lanthanide}
Ishikawa,~N.; Sugita,~M.; Ishikawa,~T.; Koshihara,~S.-y.; Kaizu,~Y. Lanthanide
  double-decker complexes functioning as magnets at the single-molecular level.
  \emph{Journal of the American Chemical Society} \textbf{2003}, \emph{125},
  8694--8695\relax
\mciteBstWouldAddEndPuncttrue
\mciteSetBstMidEndSepPunct{\mcitedefaultmidpunct}
{\mcitedefaultendpunct}{\mcitedefaultseppunct}\relax
\EndOfBibitem
\bibitem[Escalera-Moreno \latin{et~al.}(2020)Escalera-Moreno, Baldov{\'\i},
  Gaita-Ari{\~n}o, and Coronado]{escalera2020design}
Escalera-Moreno,~L.; Baldov{\'\i},~J.~J.; Gaita-Ari{\~n}o,~A.; Coronado,~E.
  Design of high-temperature f-block molecular nanomagnets through the control
  of vibration-induced spin relaxation. \emph{Chemical Science} \textbf{2020},
  \emph{11}, 1593--1598\relax
\mciteBstWouldAddEndPuncttrue
\mciteSetBstMidEndSepPunct{\mcitedefaultmidpunct}
{\mcitedefaultendpunct}{\mcitedefaultseppunct}\relax
\EndOfBibitem
\bibitem[Frisch \latin{et~al.}(2016)Frisch, \latin{et~al.} others]{Gaussian}
Frisch,~M.~J., \latin{et~al.}  Gaussian 16 Revision A.03. \emph{Gaussian 16
  Revision A.03} \textbf{2016}, \relax
\mciteBstWouldAddEndPunctfalse
\mciteSetBstMidEndSepPunct{\mcitedefaultmidpunct}
{}{\mcitedefaultseppunct}\relax
\EndOfBibitem
\bibitem[Fdez.~Galvan \latin{et~al.}(2019)Fdez.~Galvan, Vacher, Alavi, Angeli,
  Aquilante, Autschbach, Bao, Bokarev, Bogdanov, Carlson, \latin{et~al.}
  others]{fdez2019openmolcas}
Fdez.~Galvan,~I.; Vacher,~M.; Alavi,~A.; Angeli,~C.; Aquilante,~F.;
  Autschbach,~J.; Bao,~J.~J.; Bokarev,~S.~I.; Bogdanov,~N.~A.; Carlson,~R.~K.,
  \latin{et~al.}  OpenMolcas: From source code to insight. \emph{Journal of
  Chemical Theory and Computation} \textbf{2019}, \emph{15}, 5925--5964\relax
\mciteBstWouldAddEndPuncttrue
\mciteSetBstMidEndSepPunct{\mcitedefaultmidpunct}
{\mcitedefaultendpunct}{\mcitedefaultseppunct}\relax
\EndOfBibitem
\bibitem[Ungur and Chibotaru(2017)Ungur, and Chibotaru]{ungur2017ab}
Ungur,~L.; Chibotaru,~L.~F. Ab initio crystal field for lanthanides.
  \emph{Chemistry--A European Journal} \textbf{2017}, \emph{23},
  3708--3718\relax
\mciteBstWouldAddEndPuncttrue
\mciteSetBstMidEndSepPunct{\mcitedefaultmidpunct}
{\mcitedefaultendpunct}{\mcitedefaultseppunct}\relax
\EndOfBibitem
\end{mcitethebibliography}

\end{document}